\documentclass[pra,preprint,aps]{revtex4}
\usepackage{graphics}

\begin{document}

\title{Direct Determination of Harmonic and Inter-modulation distortions
 with an application to Single Mode Laser Diodes}
                              
\author{C. Tannous}
\affiliation{Laboratoire de Magn\'{e}tisme de Bretagne, UPRES A CNRS 6135, 
Universit\'{e} de Bretagne Occidentale, BP: 809 Brest CEDEX, 29285 FRANCE}

\date{March 28, 2001}


\begin{abstract}
Harmonic and Intermodulation distortions occur when a physical system is
excited with a single or several frequencies and when the 
relationship between the input and output is non-linear. Working with non-linearities in the Frequency domain is not straightforward specially when the relationship between the input and output is not trivial.
We  outline  the  complete derivation of  the  Harmonic  and
Intermodulation distortions from basic principles to a general physical system.
For illustration, the procedure is applied to the 
Single Mode laser diode where the relationship of input to output is non-trivial.
The distortions terms are extracted directly
from the Laser Diode rate equations and the method is tested 
by comparison to  many results  cited in the literature.
This methodology is general enough to be applied to the extraction of distortion
terms to any desired order in many physical systems in a general and systematic way.\\


\end{abstract}

\maketitle

\section{Introduction}
Harmonic and Intermodulation distortions occur when a signal at
a single frequency or a superposition of several signals with different frequencies propagate
in a  non-linear physical system. 
Historically it was first highlighted in the Radar and Radio Communications industry because of
the interest in understanding interference from other radars, jammers, other transmitters or modulators...
Presently, it is pervading High Energy Physics particle detectors, wideband amplifiers, satellite
communications \cite{hwang} and other areas of science and technology. To cite some of the areas of interest, it occurs in many types of devices such as Mechanical \cite {lubbock}, Acoustical (Microphones for instance), Electronic \cite {bussgang} and Microelectronic \cite{toner}, Microwave \cite{wilker}, Optical \cite{darcie}, Magnetic \cite {lim} and Superconducting \cite{monaco}.\\

Parasitic frequency terms appear either at integer multiple of the base frequency
(harmonic) or as a mixture of two or several multiples of base frequencies (intermodulation)
when several base frequencies are used (as in modulation systems for instance). These
terms can be either post-filtered or the signal can be pre-distorted in order to
avoid the appearance of these unwanted terms.\\

In this work, we introduce a general and systematic method to evaluate these
terms from the equations of motion describing the physical system at hand (Laser diode
rate equations). This case is chosen to highlight the case when the relationship
 between input and output is non-trivial.
 This paper is organised as follows. In section 2 we outline
the general procedure for expressing the output from the input and define the various distortion terms.
In section 3 we apply the procedure in detail to the Single Mode Laser diode order by order up to third and conclude in section 4. The Appendix contains the full expressions of all terms derived in section 3.

\section{General Procedure}
The Single Mode (SM) laser diode rate equations are written as:
\begin{eqnarray}
\frac{dN}{dt} & = & \frac{I}{qV}-\frac{N}{\tau_{n}}-g(N-N_{t})(1-\epsilon S)S \\
\frac{dS}{dt} & = & \frac{\Gamma \beta N}{\tau_{n}}-\frac{S}{\tau_{ph}}+\Gamma g (N-N_{t})(1-\epsilon S)S
\end{eqnarray}

$N$ represents the electron density ( $N_{t}$ at transparency) and $S$
the  photon density. $\tau_{n}$ is the electron spontaneous lifetime
and  $\tau_{ph}$  is  the  photon lifetime. $\beta$  is  the  fraction  of
spontaneous  emission coupled into the lasing  mode,  $\Gamma$  the
optical confinement factor, $g$ is the differential gain and $\epsilon$
is the gain compression parameter. $q$ is the electron charge,
$V$  the  volume  of the active region and $I$ is the  injection
current.

{\bf a- Step 1} [Elimination of $(N-N_{t})$]: \\
The input is the injection current $I$ and the output is the light intensity
represented by $S$. Since other variable such as $N$, the electron density, intervene 
in the SM equations we proceed to the elimination of that variable. When several modes
are present (or other intermediate variables) we proceed in an analogous manner
by successive elimination until we have a single equation relating input to output.\\
 
From (2) we extract the value of $(N-N_{t})$ as:
\begin{equation}
(N-N_{t})= \left[ \frac { \frac{dS}{dt} -\frac{\Gamma \beta N_{t}}{\tau_{n}}+\frac{S}{\tau_{ph}} }
               {  \Gamma g (1-\epsilon S)S + \frac{\Gamma \beta }{\tau_{n}}   } \right]
\end{equation}
                                
This can be used to find $dN/dt$ and eliminate completely $N$
from the coupled equations (1) and (2). The result is:
\begin{equation}
\frac{d}{dt} \left[ \frac { \frac{dS}{dt} -\frac{\Gamma \beta N_{t}}{\tau_{n}}+\frac{S}{\tau_{ph}} }
               {  \Gamma g (1-\epsilon S)S + \frac{\Gamma \beta }{\tau_{n}}   } \right] = 
\frac{I}{qV} -\frac{N_{t}}{\tau_{n}}  -[ g(1-\epsilon S)S + \frac{1}{\tau_{n}}] \left[ \frac { \frac{dS}{dt} -\frac{\Gamma \beta N_{t}}{\tau_{n}}+\frac{S}{\tau_{ph}} }
               {  \Gamma g (1-\epsilon S)S + \frac{\Gamma \beta }{\tau_{n}}   } \right]
\end{equation}
       
{\bf b-  Step 2}  [Small signal expansion about a static  operating point]:\\
The useful SM Laser diode regime with an injection current modulating the output light 
intensity. In the small dynamic signal case, $I=I_{0} + i$  and
$S=S_{0} + s$, where $I_{0}$ and  $S_{0}$ are the static injection current
and  photon  density  respectively, whereas, $i=i(t)$  and  $s=s(t)$, are the signals of
interest. We expand (4) [to n-th  order]:

\begin{eqnarray}
&& i=  \{ A_{1} s + B_{1} s' + C_{1} s'' \}  +  \{ A_{2} s^{2} + B_{2} s s' + C_{2} s s''
+  D_{2}[s']^{2} \} ... + \{ A_{n} s^{n}  \nonumber \\
&& \hspace{0.5cm}  +  B_{n} s^{n-1} s' + C_{n} s^{n-1} s''  + D_{n}s^{n-2} [s']^{2} \}
\end{eqnarray}

The   expansion  has  this  form  because  the  first   time
derivative is applied to a fraction whose
expansion  contains n-order terms of the form $s^{n-1} s'$  and
$s^n$. After the derivation we obtain n-order terms of the form
$s^{n-1}s'', s^{n-2}[s']^{2}$ and $s^{n-1}s'$. $s'$ and $s''$ are 
first and second time  derivatives  of s. In addition we still have  n-order
terms  of  the generic $s^{n}$ and $s' s^{n-1}$ forms coming from  the
expansion of the right hand side of (4). Hence the general n-order
 term  is $A_{n} s^{n} + B_{n} s^{n-1}s' + C_{n} s^{n-1}  s'' 
 +  D_{n}  s^{n-2} [s']^2 $ where all the parameters 
$ A_{n}, B_{n}, C_{n},  D_{n} $ depend  on the   laser  parameters. 
 The  appendix  contains   explicit expressions for some of these coefficients. 
The absence of a constant  term means that no dynamic response $i$ exists  when
there is no dynamic light excitation [$s$= 0].\\

Incidentally, we have used for simplicity of illustration the SM standard equations
that contain two coupled population $N$ (electron) and $S$ (photon) equations in the form of 
two first-order ordinary differential equations. 
The  method is in fact valid for any number of population equations  as
long  as  the  elimination  procedure  of  all  intermediate
variables  (or poulations) is  possible, leaving us with a  single  equation
relating $i$ to $s$ (Equation  (5)) or input to output.\\
The harmonic  distortions  and intermodulation  distortions  are
calculated  from the generalized transfer functions  denoted
as  $H_{n}(\omega_{1},  \omega_{2}, \omega_{3}...)$. 
They are obtained from  the  Fourier
transform  of  the  Volterra  impulse  response  $h_n$  in  the
following way:
\begin{equation}
H_{n}(j \omega_{1}, j  \omega_{2}, j \omega_{3}...)= \int_{-\infty}^{\infty}\int_{-\infty}^{\infty}...\int_{-\infty}^{\infty} h_n(t_{1},t_{2},t_{3},...)
e^{-j( \omega_{1} t_{1}+ \omega_{2} t_{2}+ \omega_{3} t_{3}...)}dt_{1}dt_{2}dt_{3}...                                          \end{equation}

For instance, the n-th order distortion is given by:
\begin{equation}
M_{n}(\pm j \omega_{1},\pm j \omega_{2},\pm j \omega_{3}...)=
20 \mbox{ log}_{10} \left\{   \frac{|H_{n}(\pm j\omega_{1}, \pm j\omega_{2}, \pm j\omega_{3}...)|}
{2^{n-1} \prod_{m=1}^{n}{|H_{1}(\pm j\omega_{m})|} }     \right\}
\end{equation} 

{\bf c- Step  3} [Method of Harmonic Input allowing the calculation
of the Volterra transfer functions $H_n$]:\\
The  Harmonic Input Method (HIM) allows us to find  directly
all the $H_{n}$'s in the following way:
\begin{enumerate}

\item Express $i$ as the sum : $exp(j\omega_{1}t)+exp(j\omega_{2}t)+exp(j\omega_{3}t)...$
\item Express $s$ as the sum:
\end{enumerate}

\begin{equation}
s= \sum_{k,l,m...=0}^{\infty}{G_{klm...}exp[j(k \omega_{1}+ l \omega_{2} + m \omega_{3}...)t]}
\end{equation}

The different $H_{n}$'s are found by direct identification of the
$G_{klm...}$ coefficients. For instance, we have:

\begin{equation}
G_{000}=   0,   G_{100}=  H_{1} (j \omega_{1}),  G_{110}=  H_{2}  (j \omega_{1},   j \omega_{2}),...
\end{equation}

\section{Order by order distortions}

Starting  from the two laser rate equations, Darcie  et  al.
\cite{darcie}  derived formulae pertaining to the second order,  third
order  and intermodulation distortions for a channel excited
by  a  superposition of two signals with angular frequencies
$\omega_{1}$ and  $\omega_{2}$.  The  second  and third order  distortions  are
calculated   respectively  at  $2\omega_{1}$  and  $3\omega_{1}$   whereas   the
intermodulation distortions are evaluated at $2\omega_{1}  -  \omega_{2}$  and
$2\omega_{2}  -  \omega_{1}$.  From  our  formula  (7),  we  can  write  these
distortions as \cite{bedrosian,mircea}:\\

a- Second order:
\begin{equation}
M_{2}( j \omega_{1}, j \omega_{1})=
20 \mbox{ log}_{10} \left\{   \frac{|H_{2}( j\omega_{1},  j\omega_{1})|}
{2 |H_{1}(j\omega_{1}) H_{1}(j\omega_{1}) |  }     \right\}
\end{equation}

b- Third order:
\begin{equation}
M_{3}( j \omega_{1}, j \omega_{1}, j \omega_{1})=
20 \mbox{ log}_{10} \left\{   \frac{|H_{3}( j\omega_{1},  j\omega_{1}, j\omega_{1})|}
{4 |H_{1}( j\omega_{1}) H_{1}( j\omega_{1}) H_{1}( j\omega_{1})|   }     \right\}
\end{equation}

c- Intermodulation:
\begin{equation}
M_{3}( j \omega_{1}, j \omega_{1}, -j \omega_{2})=
20 \mbox{ log}_{10} \left\{   \frac{|H_{3}( j\omega_{1},  j\omega_{1}, -j\omega_{2})|}
{4 |H_{1}( j\omega_{1}) H_{1}( j\omega_{1}) H_{1}( -j\omega_{2})|   }     \right\},
\end{equation}

\begin{equation}
M_{3}( j \omega_{2}, j \omega_{2}, -j \omega_{1})=
20 \mbox{ log}_{10} \left\{   \frac{|H_{3}( j\omega_{2},  j\omega_{2}, -j\omega_{1})|}
{4 |H_{1}( j\omega_{2}) H_{1}( j\omega_{2}) H_{1}( -j\omega_{1})|   }     \right\}
\end{equation}
               
The  HIM  allows us to calculate the values of  the  various
$H_{n}$'s  for n=1, 2 and 3. As an illustration of the procedure,
we  calculate $H_{1}$ and $H_{2}$ after performing steps 1 and  2  and
having  obtained  the expansion (5) to the specified  order \cite{bedrosian,mircea}.
First, we calculate $H_1$ after simply using $i=exp(j\omega t)$ in  (5)
and:
\begin{equation}
s= \sum_{k=0}^{\infty}{G_{k}exp[j(k \omega)t]}
\end{equation}
                                    
The   term   $G_{1}$  [with  the  identification  of  the   terms
multiplying $exp(j\omega t)$] obeys the relation:
\begin{equation}
1= [A_{1}+ j \omega B_{1}- \omega^{2} C_{1} ] G_{1}
\end{equation}

where  $A_{1},  B_{1} \mbox { and }C_{1}$ depend on the laser parameters. 
Using the HIM [see (9)] we can write:
\begin{equation}
H_{1}(j \omega)= G_{1}= 1/ [ A_{1}+ j \omega B_{1}- \omega^{2} C_{1} ]
\end{equation}

The  modulus of $H_{1}(j \omega)$ equal to $1/\sqrt{(A_{1}- \omega^{2} C_{1})^{2}+ 
(\omega B_{1})^{2}}$
is  what  Darcie et al. \cite{darcie} call the small-signal  frequency
response $R(\omega)$. In order to calculate second order terms,  we
use $i(t)= exp(j\omega_{1}t)+exp(j\omega_{2}t)$ in equation (5) [truncated  to
second order] along with:
\begin{equation}
s= \sum_{k,l=0}^{\infty}{G_{kl}exp[j(k \omega_{1} + l \omega_{2})t]}
\end{equation}                                 

As  stated  in  (9)  the various $G_{kl}$ are obtained  from  the
following \cite{bedrosian,mircea}: $G_{00} = 0., G_{10}= H_{1}(j \omega_{1}), G_{01}= H_{1}(j \omega_{2})  \mbox{ and } 
G_{11}=  H_{2} (j \omega_{1},   j \omega_{2})$.   Identification  of  the  terms   multiplying
$exp(j[\omega_{1}+\omega_{2}]t)$ yields:
\begin{eqnarray}
&& 0= [A_{1}+ j (\omega_{1}+\omega_{2}) B_{1}- (\omega_{1}+\omega_{2})^{2} C_{1}] G_{11}
+ [A_{2} G_{01} G_{10}+ A_{2} G_{10} G_{01} + j \omega_{1} B_{2} G_{01} G_{10} \nonumber \\
&& \hspace{1cm} + j \omega_{2} B_{2} G_{10} G_{01} -\omega_{1}^{2} C_{2} G_{01} G_{10} -\omega_{2}^{2} C_{2} G_{10} G_{01}
-  \omega_{1}\omega_{2}D_{2} G_{01} G_{10} -  \omega_{1}\omega_{2}D_{2} G_{10} G_{01}]  \nonumber \\
\end{eqnarray}

With the use of (16) this can be written as:
\begin{eqnarray}
&&H_{2} (j \omega_{1},   j \omega_{2})=  G_{11} = \nonumber \\
&&  \hspace{1mm}-[2 A_{2}+j (\omega_{1}+\omega_{2})B_{2}- (\omega_{1}^{2}+\omega_{2}^{2})C_{2} -2 \omega_{1}\omega_{2}D_{2}]
H_{1}( j\omega_{1})H_{1}( j\omega_{2})H_{1}( j [\omega_{1}+\omega_{2}])
\end{eqnarray}
The second order distortions are obtained from (10) and (20)
once the expressions of  $A_{2}, B_{2}, C_{2}, D_{2}$ are found from the
direct expansion of (4).
The  Taylor  expansion to the third order  is  made  in  the
Appendix   allowing   calculations  of   third   order   and
intermodulation distortions. Also, the values of the various
coefficients are given as functions of the laser parameters.

\section{Conclusion}

In this work, we outline a general and systematic method to evaluate Harmonic and Intermodulation distortions that occur when a signal at
a single frequency or a superposition of several signals with different frequencies propagate
in a  non-linear physical system. The method is generalisable to any order and pertains to
cases where a non-trivial relationship exists between input and output exists. \\
In the case of a SM Laser diode, we have a system of two coupled differential population equations describing the system, nevertheless a general procedure has been described in order to relate the input injection current to the output light intensity. The results obtained in this paper agree with those published in the literature and the methodology highlighted provides a self-contained framework to evaluate
distortions in a systematic way.

{\bf Acknowledgement} \\
This work started while the author was with the Department of  Electrical Engineering 
and with TRLabs in Saskatoon, Canada. The  author wishes to thank David Dodds and Michael Sieben
for introducing him to the problem. This work was supported in part by an NSERC University fellowship
grant.

%%%%%%%%%%%%%%%%%%%%%%%%%%%%%%%%%%%%%%%%%%%%%%%%%%%%%%%%%%%%%%%%%%%%%%%%%%%%
\vspace{1cm}
\centerline{\Large\bf Appendix}
\vspace{1cm}

{\bf Procedure:}\\         
In  order  to calculate the various $A_{n}, B_{n}, C_{n} \mbox{ and } D_{n}$  terms
( for n=1, 2, 3) a Taylor expansion about $S=S_{0}$ and $s'= 0$ is to be
made  to  third  order of equation (4). The  following  term
should be expanded before taking its time derivative:

\begin{equation}
\left[ \frac { \frac{dS}{dt} -\frac{\Gamma \beta N_{t}}{\tau_{n}}+\frac{S}{\tau_{ph}} }
               {  \Gamma g (1-\epsilon S)S + \frac{\Gamma \beta }{\tau_{n}}   } \right]
\end{equation}

We   can  simplify  the  expansion  considerably  using  the
following orders of magnitude valid in most single-mode laser diodes: 
 $\beta \cong 0 (\sim 10^{-3} \mbox{ to }10^{-4})$, $\tau_{n} = \infty$ ,
$1-\epsilon S_{0}$ is practically 1 and $g S_{0} \ll 1/\tau_{ph}$. 
The term becomes after these considerations:

\begin{equation}
\left[ \frac { \frac{dS}{dt} +\frac{S}{\tau_{ph}} }
               {  \Gamma g (1-\epsilon S)S   } \right]
\end{equation}

The next term to be expanded to third order is:
\begin{equation}
[g(1-\epsilon S)S + \frac{1}{\tau_{n}}] \left[ \frac { \frac{dS}{dt} -\frac{\Gamma \beta N_{t}}{\tau_{n}}+\frac{S}{\tau_{ph}} }
               {  \Gamma g (1-\epsilon S)S + \frac{\Gamma \beta }{\tau_{n}}   } \right]
\end{equation}

When  we  account for the aforementioned orders of magnitude
it becomes:
\begin{equation}
\left[ \frac { \frac{dS}{dt} + \frac{S}{\tau_{ph}} }
               {  \Gamma  } \right]
\end{equation}

The  third  order  expansion of (21) after taking  the  time
derivative gives:
\begin{eqnarray}
&& [\epsilon/(\tau_{ph}\Gamma g)]s'+ [1/(\Gamma g S_{0} )]s''+ 
[\epsilon/(\Gamma g S_{0})- 1/(\Gamma g S_{0}^{2})]([s']^{2}+
ss'')  \nonumber \\
&&   \hspace{0.5cm} + [\epsilon^{2}/(\Gamma g S_{0}) - (3 \epsilon S_{0}-2 \epsilon^{2}S_{0}^{2}
-1)/(\Gamma g S_{0}^{3})](s^{2}s''+2s[s']^{2})
\end{eqnarray}

Adding the contributions from (23) $[s'/ \Gamma + s/(\tau_{ph}\Gamma )]$
 and  term by term identification gives for the various orders:\\

{\bf First order:} $A_{1} s + B_{1}s' + C_{1}s''$ where:
\begin{eqnarray}
A_{1} & =  & 1/(\tau_{ph}\Gamma) \nonumber \\
B_{1} & = &  \epsilon /(\tau_{ph}\Gamma g )+1/\Gamma \nonumber \\
C_{1} & = & 1/(\Gamma g S_{0})
\end{eqnarray}

{\bf Second order:} $A_{2} s^{2} + B_{2} s s' + C_{2} s s'' + D_{2}[s']^{2}$ with:
\begin{eqnarray}
A_{2} & =  & 0  \nonumber \\
B_{2} & =&  0  \nonumber \\
C_{2} & = & [\epsilon/(\Gamma g S_{0})- 1/(\Gamma g S_{0}^{2})]  \nonumber \\
D_{2} & = & C_{2} 
\end{eqnarray}

{\bf Third order:} $A_{3} s^{3} + B_{3} s^{2} s' + C_{3} s^{2} s'' + D_{3}s[s']^{2}$ with:
\begin{eqnarray}
A_{3} & =  & 0  \nonumber \\
B_{3} & =&  0  \nonumber \\
C_{3} & = & [\epsilon^{2}/(\Gamma g S_{0}) - (3 \epsilon S_{0}-2 \epsilon^{2}S_{0}^{2} -1)/(\Gamma g S_{0}^{3})]  \nonumber \\
D_{3} & = & 2C_{3} 
\end{eqnarray}

The frequency response:
\begin{equation}
|H_{1}(j \omega)| = 1/ \left[ A_{1} \sqrt{(1- \omega^{2} C_{1}/A_{1})^{2}+ 
(\omega B_{1}/A_{1})^{2}} \right]
\end{equation}

and  the  second  harmonic  distortion  given  by  (10)  and
transformed with the help of (19):

\begin{equation}
M_{2}( j \omega_{1}, j \omega_{1})=
20 \mbox{ log}_{10} \{ 2 \omega_{1}^{2}|C_{2} H_{1}(2j \omega_{1})| \}
\end{equation}

agree with the results of Darcie et al \cite{darcie}.\\
In  order  to  calculate  third  order  and  intermodulation
effects,  we  use eq. (5) truncated to third order  and  the
values of the $ A_{n}, B_{n}, C_{n}$ coefficients (n=1, 2, 3):
\begin{equation}
i=  [A_{1} s + B_{1} s' + C_{1} s''] + C_{2}[s s'' + [s']^{2}] + C_{3}[s^{2}s'' + 2s[s']^{2}]        
\end{equation}

Apply   the   HIM   to   the   above   equation   with:  
 $i= exp(j \omega_{1}t)+exp(j \omega_{2}t)+exp(j \omega_{3}t)$  and:
\begin{equation}
s= \sum_{k,l,m=0}^{\infty}{G_{klm}exp[j(k \omega_{1} + l \omega_{2} + m \omega_{3})t]}
\end{equation}

Proceeding   as  before  (eq.18)  we  calculate   the   $G_{klm}$
coefficients  as  well  as  the  various  Volterra  transfer
functions: 
\begin{eqnarray} 
&& G_{000}= 0, G_{100}= H_{1}(j \omega_{1}), G_{010}= H_{1}(j \omega_{2}),
G_{001}= H_{1}(j \omega_{3}), \nonumber \\
&& G_{110}= H_{2}(j \omega_{1},j \omega_{2}),G_{101}= H_{2}(j \omega_{1},j \omega_{3})  \nonumber \\
&& G_{011}= H_{2}(j \omega_{2},j \omega_{3})  \hspace{0.5cm} \mbox{ and finally } G_{111}= H_{3}(j \omega_{1},j \omega_{2},j \omega_{3})
\end{eqnarray}

from eq.(30).   \\  
Collecting  the  terms  multiplying  $exp[j(\omega_{1} + \omega_{2} +  \omega_{3})t]$ gives:
\begin{eqnarray}
&& 0=[A_{1}+j (\omega_{1} + \omega_{2} +  \omega_{3}) B_{1}
- (\omega_{1} + \omega_{2} +  \omega_{3})^{2} C_{1}]G_{111} \nonumber \\
&& -2 C_{2} (\omega_{1} + \omega_{2} +  \omega_{3})^{2}[ 2 G_{110} G_{001}
+ 2 G_{101} G_{010} + 2 G_{100} G_{011} ]  \nonumber \\
&& -3 C_{3} (\omega_{1} + \omega_{2} +  \omega_{3})^{2} [ 6 G_{100} G_{010}  G_{001}]
\end{eqnarray}

Using the above relations between the  $G_{klm}$ and the $ H_{n}$ 's and
(16) we get:
\begin{eqnarray}
H_{3}(j \omega_{1},j \omega_{2},j \omega_{3} )= (\omega_{1} + \omega_{2} +  \omega_{3})^{2} H_{1}(j [\omega_{1}+ \omega_{2}+  \omega_{3}]) [4 C_{2} \{H_{2}(j \omega_{1},j \omega_{2}) H_{1}(j \omega_{3}) \nonumber \\
 + H_{2}(j \omega_{1},j \omega_{3}) H_{1}(j \omega_{2}) + H_{2}(j \omega_{2},j \omega_{3}) H_{1}(j \omega_{1}) \}  \nonumber \\
 + 18  C_{3} H_{1}(j \omega_{1})H_{1}(j \omega_{2})H_{1}(j \omega_{3})] \nonumber \\
\end{eqnarray}

Using  expressions (16) for $ H_{1}(j \omega) $ and (19)  for 
$ H_{2}(j \omega_{1},j \omega_{2})$   along with eq.(11), (12) and (13) the third order  and
intermodulation distortions can be evaluated and they  agree
again with Darcie et al. \cite{darcie} results.

\end{document}